\documentclass[fleqn,10pt]{wlscirep}
\usepackage[utf8]{inputenc}
\usepackage[T1]{fontenc}
\title{The Duel of Magnetic Interactions \& Structural Instabilities: Itinerant Frustration in the Triangular Lattice Compound LiCrSe$_2$}

\author[1,*]{E.~Nocerino}
\author[2]{S.~Kobayashi}
\author[3,4]{C.~Witteveen}
\author[5]{O.~K.~Forslund}
\author[1]{N. Matsubara}
\author[6]{C.~Tang}
\author[7]{T.~Matsukawa}
\author[7]{A.~Hoshikawa}
\author[8,9]{A.~Koda}
\author[10]{K.~Yoshimura}
\author[11]{I.~Umegaki}
\author[5]{Y.~Sassa}
\author[3]{F.~O.~von~Rohr}
\author[12]{V.~Pomjakushin}
\author[13,14]{J.~H.~Brewer}
\author[15,16]{J.~Sugiyama}
\author[1,+]{M.~M{\aa}nsson}

\affil[1]{KTH Royal Institute of Technology, Department of Applied Physics, Alba Nova University Center, Stockholm, SE-114 21, Sweden}

\affil[2]{Japan Synchrotron Radiation Research Institute (JASRI), 1-1-1 Kouto, Sayo 679-5198, Japan}

\affil[3]{Department of Quantum Matter Physics, University of Geneva, 24 Quai Ernest-Ansermet,
1211 Geneva 4, Switzerland}
\affil[4]{Department of Physics, University of Zürich, Winterthurerstr. 190, 8057 Zürich, Switzerland}

\affil[5]{Chalmers University of Technology, Department of Physics, G$\ddot{o}$teborg, SE-412 96, Sweden}

\affil[6]{Diamond House, Harwell Science and Innovation Campus, Fermi Ave, Didcot OX11 0DE, UK}

\affil[7]{Frontier Research Center for Applied Atomic Sciences, Ibaraki University, 162-1 Shirakata, Tokai, Ibaraki 319-1106, Japan}

\affil[8]{Organization (KEK), Tsukuba, Ibaraki 305-0801, Japan}

\affil[9]{Department of Materials Structure Science, The Graduate University for Advanced Studies, Tsukuba, Ibaraki 305-0801, Japan}

\affil[10]{Department of Chemistry, Graduate School of Science, Kyoto University, Kyoto 606-8502, Japan}

\affil[11]{Muon Science Laboratory, Institute of Materials Structure Science, KEK, Tokai, Ibaraki 319-1106, Japan}

\affil[12]{Laboratory for Neutron Scattering and Imaging, Paul Scherrer Institute, 5232 Villigen PSI, Switzerland}

\affil[13]{Department of Physics and Astronomy, University of British Columbia, Vancouver, British Columbia, Canada V6T 1Z1}

\affil[14]{TRIUMF, 4004 Wesbrook Mall, Vancouver, British Columbia, Canada V6T 2A3}

\affil[15]{Neutron Science and Technology Center, Comprehensive Research Organization for Science and Society (CROSS), Tokai, Ibaraki 319-1106, Japan}

\affil[16]{Advanced Science Research Center, Japan Atomic Energy Agency, Tokai, Ibaraki 319-1195, Japan}

\affil[*]{nocerino@kth.se}

\affil[+]{condmat@kth.se}


\begin{abstract}
The recent synthesis of the chromium selenide compound LiCrSe$_2$ constitutes a valuable addition to the ensemble of two-dimensional triangular lattice antiferromagnets (2D-TLA). In this work we present the very first comprehensive study of the combined low temperature nuclear and magnetic structure established in this material. Details on the connection between Li-ion dynamics and structural changes are also presented along with a direct link between atomic structure and spin order via a strong magnetoelastic coupling. LiCrSe$_2$ was found to undergo a first order structural transition from a trigonal crystal system with space group $P\bar{3}m1$ to a monoclinic one with space group $C2/m$ at $T_{\rm s}=30$~K. Such restructuring of the lattice is accompanied by a magnetic transition at $T_{\rm N}=30$~K, with the formation of a complex spin arrangement for the Cr$^{3+}$ moments. Refinement of the magnetic structure with neutron diffraction data and complementary muon spin rotation analysis reveal the presence of two incommensurate magnetic domains with a up-up-down-down arrangement of the spins with ferromagnetic (FM) double chains coupled antiferromagnetically (AFM). In addition to this unusual arrangement, the spin axial vector is modulated both in direction and modulus, resulting in a spin density wave-like order with periodic suppression of the Cr moment along the chains. This behavior is believed to appear as a result of strong competition between direct exchange AFM and superexchange FM couplings established between both nearest neighbor and next nearest neighbor Cr$^{3+}$ ions. We finally conjecture that the resulting magnetic order is stabilized via subtle vacancy/charge order within the Li layers, potentially causing a mix of two co-existing magnetic phases within the sample.
\end{abstract}
\begin{document}

\flushbottom
\maketitle

\thispagestyle{empty}

\section*{Introduction}

In periodic crystalline lattices containing magnetic atoms, long range ordering of their electronic magnetic moments can be stabilized by the exchange couplings among them at low temperatures. In some systems however, due to peculiar structural properties, not all such interactions can be satisfied simultaneously, leading to unconventional magnetic ground states, which possess unusual physical properties. One example of such systems is provided by geometrically frustrated triangular lattice antiferromagnets (TLA). Their crystal structure consists of two-dimensional triangular lattice (2D-TL) layers, in which each corner of the triangular lattice is occupied by a magnetic atom, with nearest neighbor antiferromagnetic (AFM) interactions that cannot be stabilized \cite{Anderson_1973}. Some prominent examples of 2D-TLA are found amid transition metal oxides (TMO), such as NaCoO$_2$ \cite{Sugiyama_2003,Hertz_2008,Schulze_2008,Medarde_2013,Sassa_2018} or Li/Na/KCrO$_2$ \cite{Soubeyroux_1979,Sugiyama_2009}. Among the currently available chromium based 2D-TLA, compounds of the type $A$Cr$X_2$, where $A$ is a monovalent atom and $X$ is a chalcogen element, have been extensively investigated as geometrically frustrated $S=3/2$ Heisenberg spin systems. The interest around these materials lies in the fact that they manifest an extremely luscious smörgåsbord of complex magnetic ground states \cite{rasch2009magnetoelastic,carlsson2011suppression,lopes2011local,ji2009spin,van1971preparation,engelsman1973crystal,damay2011magnetoelastic,van1973magnetic} and attractive technological properties (e.g. multiferroicity) \cite{Hartmann_2013,Halley_2014,Soda_2016}. These systems entail higher energy spin configurations, therefore, in the low temperature regime, often a compromise between spin and lattice degrees of freedom has to be reached. The compromise can be realized through a first-order energetically beneficial distortion of the lattice leading to a new symmetry of the crystal, in which new interaction paths are realized and a magnetic order can be stabilized. When such an interplay between strongly correlated spins and geometrical degrees of freedom is realized, it is said that the system manifests magnetoelastic coupling. The appeal for materials undergoing these kind of phenomena is twofold, as they constitute a unique playground for fundamental scientific research \cite{rasch2009magnetoelastic,carlsson2011suppression,lopes2011local,ji2009spin,van1971preparation,engelsman1973crystal,damay2011magnetoelastic,van1973magnetic,Bazazzadeh_2021}, but also lend themselves to potential novel technical applications \cite{Grimes_2011,G_Saiz_2022}.

The recent synthesis of LiCrSe$_2$, via Li intercalation in CrSe$_2$ by Kobayashi $et~al.$ \cite{kobayashi2014successive,kobayashi2016competition}, constitutes a valuable addition to the ensemble of TLAs that manifest magnetoelastic coupling since, probably due to the difficulty in manufacturing alkali chromium selenides, the most of the studies in this framework are performed on oxides and sulfides. Beyond the fundamental physics interest surrounding LiCrSe$_2$, recent theoretical works foresee remarkable electronic properties, such as the presence of tunable field-free topological spin textures, which make this material appealing for applications in spintronics circuitry \cite{li2022large,xu2020intrinsic,kumari2021recent}. The room temperature crystal structure of LiCrSe$_2$ was identified as trigonal by synchrotron X-ray diffraction, with space group $P\bar{3}m1$. Here, the Cr$^{3+}$ ions create 2DTL layers, stacked along the $c$ axis, by the connection of edge-sharing CrSe$_6$ octahedra in the $ab-$plane \cite{kobayashi2019linear}. Indications of a first order transition towards a lower symmetry crystal system were observed around $T=30$~K with in-house X-ray diffraction \cite{kobayashi2016competition}; here it is suggested a transition towards a monoclinic structure. From bulk magnetic characterization methods, LiCrSe$_2$ was found to undergo a first order-like magnetic transition, with an AFM character, also around $T_{\rm N}=30$~K, which is a hint of magnetoelastic coupling. This behavior is similar to the one observed in the magnetoelastic TLA-Cr sulfides AuCrS$_2$ and AgCrS$_2$ \cite{carlsson2011suppression,damay2011magnetoelastic}. However, remarkable qualitative differences in the structural distortions and magnetic properties between LiCrSe$_2$ and the sulfides, indicate that LiCrSe$_2$ would select a different magnetic ground state with respect to the one realized in AuCrS$_2$ and AgCrS$_2$. Indeed, the small negative value of the LiCrSe$_2$ Curie-Weiss temperature $\Theta=-28$~K and the small frustration index $f = \frac{|\Theta|}{T_{\rm N}}\approx1$ estimated in reference \cite{kobayashi2016competition}, are suggestive of a weak nearest neighbor magnetic interaction (weaker than the one found in the sulfides) in favor of a strong further neighbor magnetic interaction (stronger than the one found in the sulfides). These results indicate that the further neighbor interactions have a leading role in the determination of the magnetic ground state in LiCrSe$_2$, making it a promising candidate for the realization of unconventional magnetic structures.

In this work we investigated the temperature evolution of the crystal and magnetic structure of polycrystalline LiCrSe$_2$ by means of neutron diffraction (ND), synchrotron X-ray diffraction (XRD) and muon spin rotation ($\mu^+$SR). Our observations confirm the conjectures of Kobayashi and coworkers, as a complex magnetic structure seems to be established in the sample, with two incommensurate magnetic domains with periodic modulation in direction and modulus of the Cr$^{3+}$ spin axial vector. The two domains contribute differently to the neutron diffraction pattern, suggesting that one is larger than the other, as also indicated by the $\mu^+$SR analysis. A solution for the main and secondary magnetic structures is proposed, together with a possible interpretation of such a spin ordering in terms of the magnetic exchange interactions for the main phase.

\section*{Results}

The experimental results and data analysis are presented below in the respective sub-sections. First the low-temperature $\mu^+$SR investigation is  presented, followed by the NPD study and complementary synchrotron XRD measurements as a function of temperature.

\begin{figure*}[ht]
  \begin{center}
    \includegraphics[keepaspectratio=true,width=177 mm]{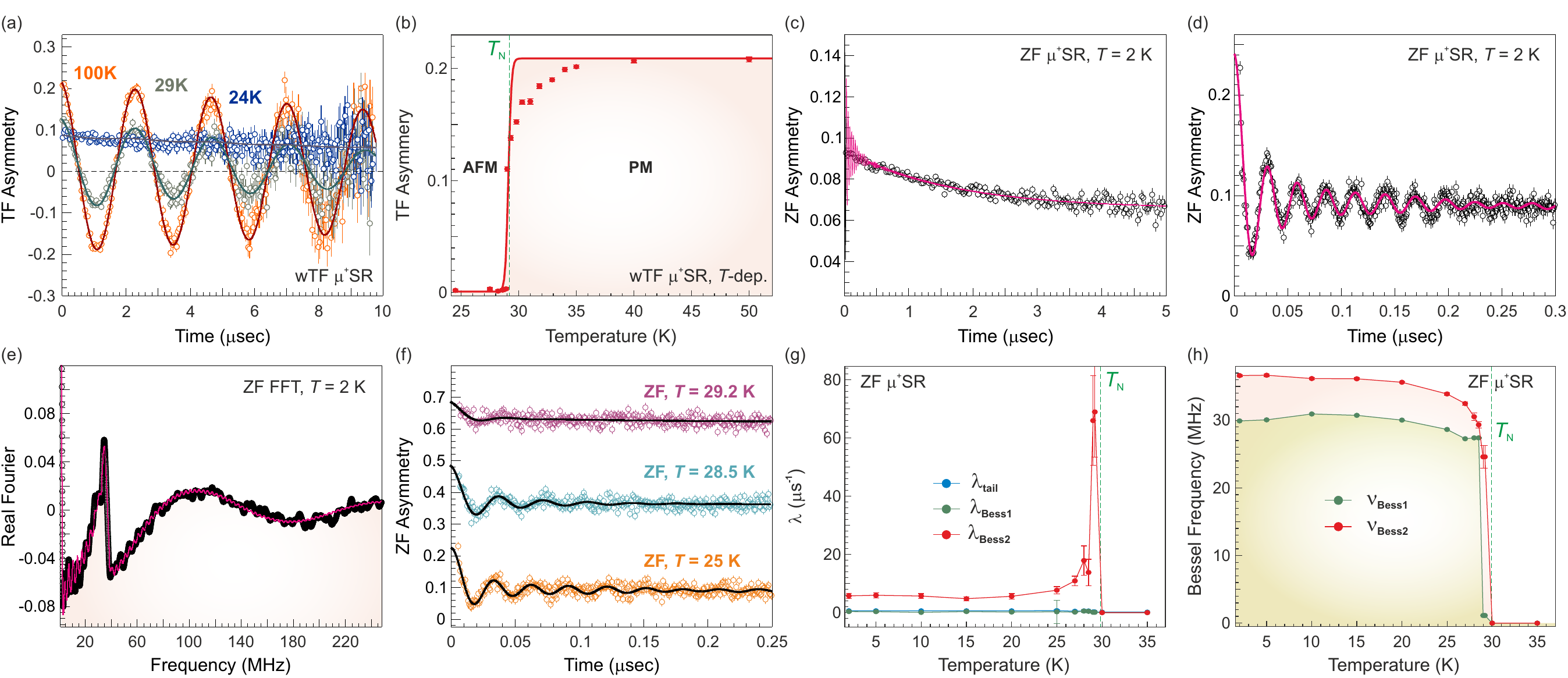}
  \end{center}
  \caption{(a) Weak transverse field (wTF) $\mu^+$SR time spectra at selected temperatures with best fit to Eq.~\ref{wTFit} shown as solid lines. (b) Temperature dependence of the wTF asymmetry with the continuous red line representing a fit to a sigmoid function giving the AFM transition temperature $T_{\rm N}=29.11\pm0.06$~K. (c,d) Zero-field (ZF) $\mu^+$SR time spectrum at $T=2$~K with the fit to Eq.~\ref{eqmuon} (solid magenta line) shown in (c) long and (d) short time domain, respectively. (e) Fast Fourier Transform (FFT) of the spectrum shown in (d). (f) ZF muon time spectra at selected temperatures $T=25$~K, 28.5~K and 29.2~K in the short time domain. The spectra are shifted along the y-axis for clarity of display. The black continuous lines are fits to Eq.~\ref{eqmuon}. (g,h) Temperature dependencies of the ZF fit parameters extracted from Eq.~\ref{eqmuon} for the two oscillating components of the signal plus the slow tail component.}
  \label{muon}
\end{figure*}

\subsection*{\label{nopres}$\mu^+$SR Results}

A zero-field (ZF) and weak transverse field (wTF) bulk $\mu^+$SR experiment has been performed for different temperatures, in order to observe the evolution of the sample's internal magnetic field distribution. In the wTF mode an external magnetic field of weak magnitude is applied (here, wTF~=~30~G), with its flux lines perpendicular to the initial direction of the muon spin polarization. In this set-up it is possible to follow the magnetic phase transition, while lowering the temperature of the system, as the muon spin precession (Larmor precession) is more and more affected by the sample's strong internal field, induced by long range magnetic ordering, at the expense of the externally applied weak field (for more details about the $\mu^+$SR experimental technique refer to \cite{blundell1999spin}). The wTF time spectra are well fitted using the following function:
\begin{eqnarray}
 A_0 \, P_{\rm TF}(t) \; = \; A_{\rm TF}\cos(2\pi \nu_{\rm TF}t \; + \; \frac{\pi \phi}{180})\cdot{}e^{(-\lambda_{\rm TF} t)} \; + \; A_{\rm tail}\cdot{}e^{(-\lambda_{\rm tail} t)}
\label{wTFit}
\end{eqnarray}
were $A_0$ is the initial asymmetry at time zero, $P_{\rm TF}$(t) is the muon spin polarization function, $A_{\rm TF}$ is the oscillating asymmetry while $A_{\rm tail}$ is the asymmetry of a slow relaxing exponential term due to the components of the internal field which are parallel to the initial polarization direction of the muons spins. This "tail" component, is also sometimes called the "powder average" signal when measuring polycrystalline (powder) samples, which then ideally constitute 1/3 of the total asymmetry. Further, $\nu_{\rm TF}$ is the muon spin precession frequency having the phase $\phi$ (which in this case has a value of $\approx$ 12$^{\circ}$ in the entire temperature range), and finally $\lambda_{\rm TF}$ and $\lambda_{\rm tail}$ are the depolarisation rates for the respective polarization components. wTF $\mu^+$SR time spectra at selected temperatures are shown in Fig.~\ref{muon}(a) with best fits to Eq.~\ref{wTFit} as solid lines. Further fitting of the full wTF temperature dependence and extracting the wTF asymmetry as a function of temperature ($A_{\rm TF}(T)$) allow us to estimate the magnetic transition temperature of LiCrSe$_2$. The result is shown in Fig.~\ref{muon}(b) where $A_{\rm TF}(T)$ is fitted to a sigmoid function giving $T_{\rm N}=29.11\pm0.06$~K, which is in very good agreement with previously published bulk characterizations \cite{kobayashi2016competition}. From the wTF data it is also possible to deduce that more or less 100\% of the sample enters the magnetically ordered state. Further, and contrary to our previous $\mu^+$SR studies of the related LiCrTe$_2$ compound \cite{Nocerino_2022}, we do not seem to have any clear missing asymmetry fraction above $T_{\rm N}$, i.e. absence of any evident FM impurities in the current LiCrSe$_2$ sample.

In the ZF configuration, the muon spin only senses the local internal magnetic field of the sample and if long-range magnetic order is established, a clear precession signal (oscillation) will appear. The ZF $\mu^+$SR time spectrum at base temperature ($T=2$~K) in LiCrSe$_2$ is displayed in Fig.~\ref{muon}(c,d).

At a first look, the ZF signal in the short time domain [Fig.~\ref{muon}(d)] presents very clear oscillations, indicating the occurrence of AFM spin order. However, initial fitting attempts, using two simple exponentially relaxing cosine functions, gave poor results. Several problems were found, e.g., the full asymmetry could not be recovered, the fit could only work if the phase of the cosine functions would be left free and acquire very large un-physical values, and the chosen model could not fully capture the features of the Fast Fourier transformed (FFT) signal [Fig.~\ref{muon}(e)]. After testing several combinations of different oscillatory functions, the best fit to the data was achieved with a combination of two spherical Bessel functions $j_0$:
\begin{eqnarray}
 A_0 \, P_{\rm ZF}(t) \; =&
A_{\rm Bess1}j_0(2\pi \nu_{\rm Bess1}t \; + \; \frac{\pi \phi_{\rm Bess1}}{180})\cdot{}e^{(-\lambda_{\rm Bess1} t)} \; + \; A_{\rm Bess2}j_0(2\pi \nu_{\rm Bess2}t \; + \; \frac{\pi \phi_{\rm Bess2}}{180})\cdot{}e^{(-\lambda_{\rm Bess2} t)} \; +
\cr
 &+ \; A_{\rm tail}\cdot{}e^{(-\lambda_{\rm tail} t)} \; + \; A_{\rm BG}\cdot{}e^{(-\lambda_{\rm BG} t)}.
\label{eqmuon}
\end{eqnarray}
Here $A_0$ is the initial asymmetry, $P_{\rm TF}$(t) is the muon spin polarization function, $A_{\rm Bess1}$ and $A_{\rm Bess2}$ are the asymmetries of the oscillating part of the signal, $A_{\rm BG}$ is a small background component possibly coming from muons stopping in the sample holder, and $A_{\rm tail}$ is the asymmetry of the tail component. The frequency of the Larmor precession is given by $2\pi\cdot\nu_{\rm Bess}$, and $\phi$ is the relative phase of the oscillating signals. Here $\phi\approx0$ for both the Bessel functions in the entire temperature range. Finally, $\lambda_{\rm Bess}$, $\lambda_{\rm tail}$ and $\lambda_{\rm BG}=0$ are the depolarisation rates for the respective polarization components. Figure~\ref{muon}(d,f) shows the ZF raw data in the short time domain, for different temperatures, with the fitting curves as solid lines. As seen from these results, as well as the accurate fit to the FFT at $T=2$~K in a large frequency range in Fig.~\ref{muon}(e), Eq.~\ref{eqmuon} provides a good model to the experimental data. 

The temperature dependence of the relevant ZF fitting parameters are shown in Fig~\ref{muon}(g,h). For both the relaxation rates ($\lambda$) as well as oscillation frequencies ($\nu_{\rm Bess}$) it is clear that an abrupt, first order like, transition occurs for $T_{\rm N}\approx30$~K. In more detail, the depolarization rate, and especially $\lambda_{\rm Bess2}$ display a very typical and strong increase just below $T_{\rm N}$, indicating the onset of spin dynamics/fluctuations as the magnetic phase transition approaches. The trend for $\lambda_{\rm Bess1}$ is similar, however, the absolute values are much smaller. Here it is also important to emphasize that $A_{\rm Bess2}\approx12\cdot A_{\rm Bess1}$, i.e. have more than an order of magnitude larger volume fraction. What such differences means is hard to rigorously deduce using only the $\mu^+$SR data. Nevertheless, it indicates that the two oscillating components originate from muons experiencing two rather different local magnetic environments (this point will be further discussed below). Finally, the temperature dependence of the oscillation frequencies are shown in Fig.~\ref{muon}(h), which reflect the magnetic order parameter of the phase transition. The behavior of the two components is in fact very similar, the oscillation frequencies have similar values, and follow the same temperature dependence. The latter shows an approximately constant trend below $T_{\rm N}$, with a slow downturn of both $\nu_{\rm Bess1}$ and $\nu_{\rm Bess1}$. This is then abruptly interrupted at $T_{\rm N}\approx30$~K which, we know from previous studies \cite{kobayashi2016competition}, coincides with a structural phase transition at $T_{\rm s}$.

The presence of large non-zero phases ($\phi$) for cosine functions and replacement with a damped $j_0$ Bessel is commonly used in the $\mu^+$SR community to model incommensurate magnetic structures. This solution has been well established and proven, both theoretically and experimentally, to well describe incommensurate spin density waves associated to single $q-$vectors \cite{major1986zero,amato1995muon,Andreica_Thesis}. The presence of two distinct Bessel functions with two frequencies of similar value in LiCrSe$_2$ could be due to the presence of two different muon stopping sites in the lattice, thereby probing the same incommensurate magnetic structure in different fashion. Another option would be that we have two slightly different magnetic phases present in the sample. Additional discussions on this matter will follow below.

\begin{figure*}[ht]
\centering
  \includegraphics[scale=0.78]{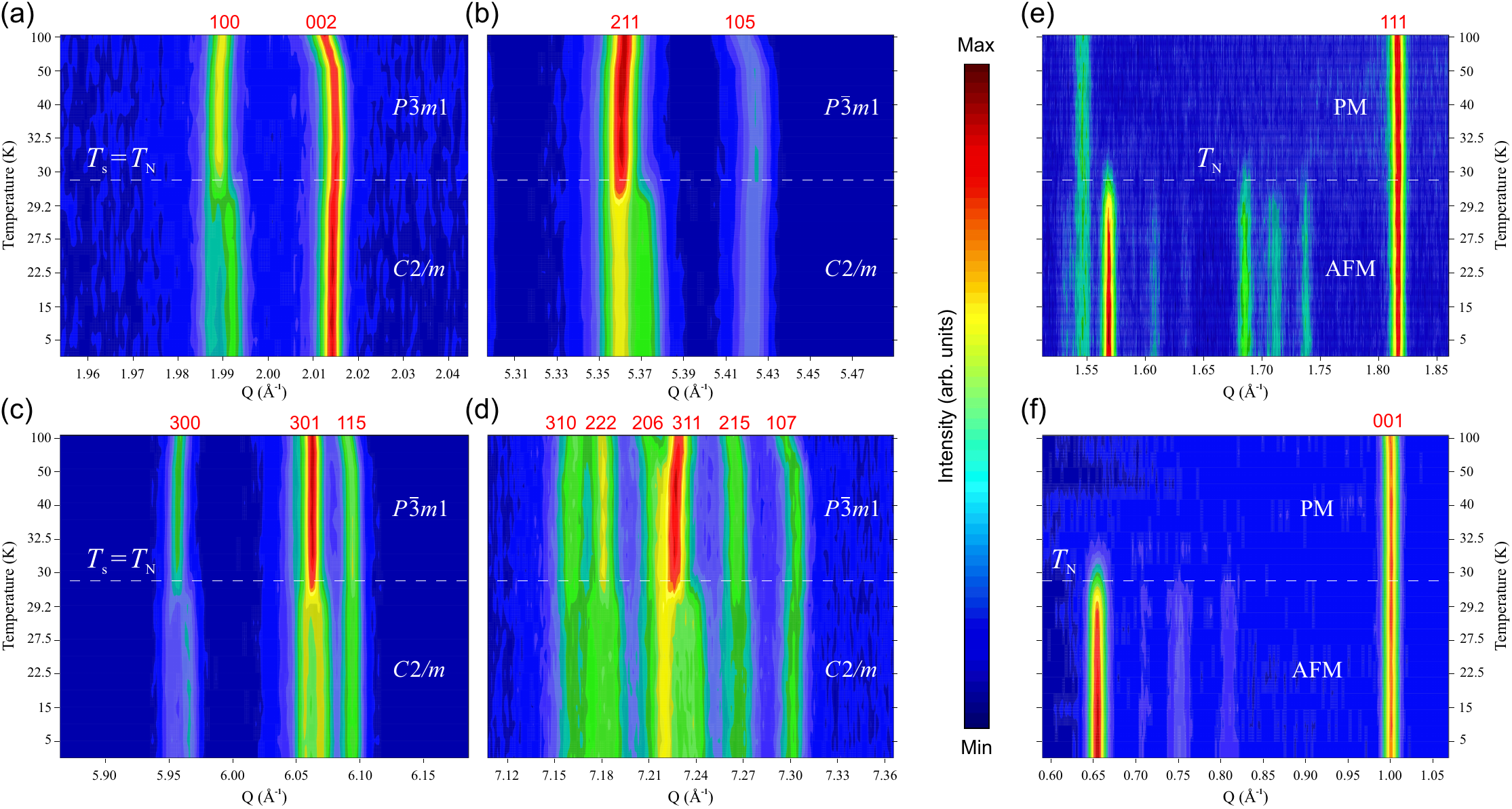}
  \caption{2D color maps showing the temperature evolution of the NPD data. (a-d) The nuclear Bragg peak splitting in different $Q-$ranges below $T_{\rm s}\approx30$~K indicates the structural phase transition from $P\bar{3}m1$ to $C2/m$. (e,f) Magnetic Bragg peaks, plotted together with some (non splitted) nuclear peaks, appear below $T_{\rm N}\approx30$~K. Please note the nonlinear temperature scale (y-axis), having a higher density of points in proximity of the transition, and that $T_{\rm s}=T_{\rm N}$. The Miller indices of the reflections in the high temperature structure are reported in red for clarity of display.}
  \label{split}
\end{figure*}

\subsection*{XRD and NPD Results: Structural Refinement}
\begin{figure*}[ht]
\centering
  \includegraphics[keepaspectratio=true,width=175 mm]{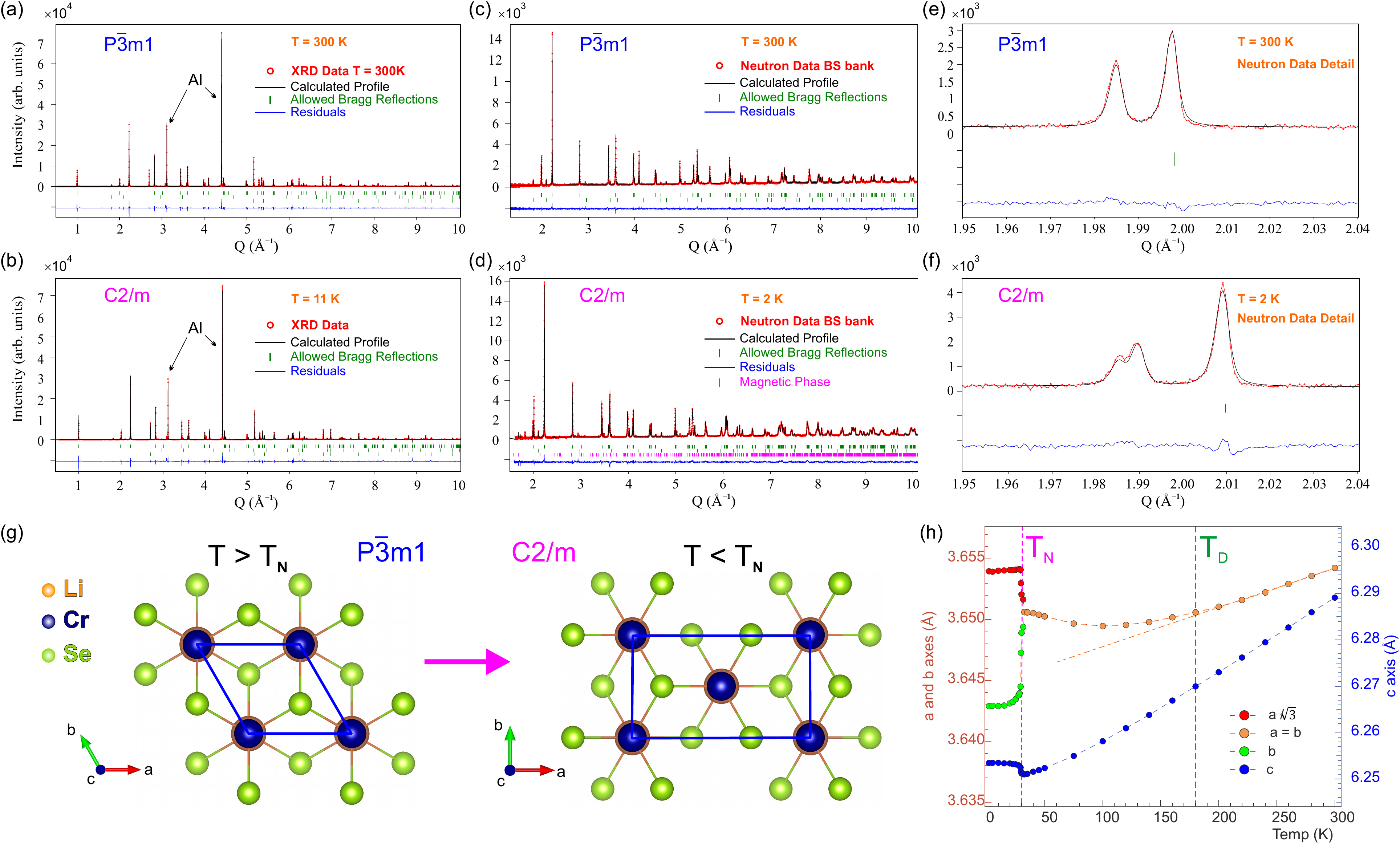}
  \caption{ a,b) X-ray diffraction patterns with the corresponding calculated patterns resulting from structural refinement at room and base temperature. c-f) Neutron diffraction patterns (detector bank BS) with the corresponding calculated patterns resulting from structural refinement at room and base temperature. The peak splitting is clearly visible in the close-up plot in figure f. g) Graphic representation of the crystal structure in polycrystalline LiCrSe$_2$ at 300 K and 2 K oriented along the $c$ axis for clear visualization of the $ab-$plane. h) Temperature dependence of the lattice parameters in LiCrSe$_2$. The y axis of the lattice parameter $c$ is plotted on the right for clarity of display. The values of the lattice parameters below the transition are reported in the hexagonal setting for correct comparison with the higher temperature values.}
  \label{struc}
\end{figure*}

The high and low temperature nuclear structure of LiCrSe$_2$, was determined from high-resolution neutron powder diffraction (NPD) studies [see Fig.~\ref{split}(a-f) and Fig.~\ref{struc}(c-f)] along with complementary synchrotron X-ray diffraction (XRD) data [see Fig.~\ref{struc}(a,b)]. The refinement of the diffraction patterns was carried out by means of the Rietveld refinement method. A global fit was employed, which included all the detector banks while keeping the cell parameters, the atomic positions and the basis vectors of the magnetic moment as common parameters. The calculated NPD/XRD pattern consists of a majority LiCrSe$_2$ phase plus a very tiny Li$_2$Se non mangetic impurity phase ($<5\%$, consistent with the $\mu^+$SR data). For the XRD an additional Al phase is included to account for the sample holder/capillary. Further, for NPD the magnetic phase is present below $T_{\rm N}$ in the low $Q$-range, while disregarded for large Q due to the magnetic form factor \cite{Lovesey_1986} (by means of the command MDLIM in FullProf).

The room temperature crystal structure [see Fig.~\ref{struc}(g)] was found to be well modeled  by the trigonal $P\bar{3}m1$ space group (no. 164), with lattice parameters $a=b=3.65425(3)$~Å, $c=6.28911(4)$~Å, $\alpha=\beta=90^{\circ}$, $\gamma=120^{\circ}$, in which Cr$^{3+}$ ions occupy the vertices of the hexagonal unit cell, hereby creating the 2D-TLA. Below $T\approx30$~K~$=T_{\rm s}=T_{\rm N}$, the diffraction pattern undergoes a dramatic change, as clearly displayed in Fig.~\ref{split}. Several nuclear Bragg peaks split into double peaks [Fig.~\ref{split}(a-d)] and, at the same time, the magnetic Bragg peaks appear [Fig.~\ref{split}(e,f)]. First, we focus on the structural change. The increased number of Bragg reflections in the diffraction pattern, due to the peak splitting, is suggestive of a symmetry loss in the system, possibly owing to the disappearance of the 3-fold symmetry axis. Moreover, no clear super-lattice reflection peaks seem to appear in the diffraction pattern. Under the light of these observations, and from a group/subgroup relationship argument, the most probable final space group for such a structural transition would be the monoclinic $C2/m$ (no. 12). This is because $C2/m$ is the only monoclinic maximal subgroup of the parent space group $P\bar{3}m1$ with a single step in its symmetry reduction path (see Bärnighausen tree for $P\bar{3}m1$ from the Bilbao Crystallographic Server \cite{ivantchev2000subgroupgraph}). Indeed, Rietveld refinement of the low temperature diffraction pattern with $C2/m$ provided a good agreement with the experimental data, as it was able to correctly index all the additional new peaks [Fig.~\ref{struc}(b,d,f)]. Nevertheless, in the light of the first-order nature of the structural transition, the possibility that the group subgroup relationship might not hold in this particular case must be considered. Several other monoclinic space groups were tested with a trial an error approach; among them $Cm$ (no. 8) provided the best matching between the calculated and the observed pattern, which was however comparable to $C2/m$. The fact that the loss of the 2-fold symmetry axis from $C2/m$ to $Cm$ did not improve the refinement, made $C2/m$ a more reasonable choice.
This space group was also suggested from previous in-house XRD measurements \cite{kobayashi2016competition}, and we have now confirmed such suggestion with high-resolution experimental methods. The base temperature ($T=2$~K) crystal structure was therefore modeled with the monoclinic $C2/m$ space group, with lattice parameters $a$ = 6.32904(8) Å, $b$ = 3.64287(5) Å, $c$ = 6.25345(7) Å, $\alpha$ = $\gamma$ = 90$^{\circ}$, and $\beta$ = 90.08(13)$^{\circ}$. Here the Cr$^{3+}$ ions occupy the vertices of a unit cell, which is nearly rectangular in the $ab-$plane, with one additional Cr atom in the central point (0.5, 0.5, z) [see Fig.~\ref{struc}(g)].
The detailed result of the refinement is displayed in Fig.~\ref{struc}(e,f), with a plot of the diffraction patterns (observed and calculated) at high and low temperature, in the high resolution BS detector bank. It is important to emphasize that in this bank (and $Q-$range) any potential low-temperature magnetic peak is virtually invisible, and does not interfere with the structural refinement. The complementary XRD patterns are also shown [Fig.~\ref{struc}(a-b)], and the structural refinement for XRD/NPD measurements is fully consistent, further allowing us to effectively discern between nuclear and magnetic contributions. It should also be noted that refinement of fractional occupancies did not improve the goodness of the fit, which allow us to conclude that the LiCrSe$_2$ sample has a stoichiometric composition very close to the ideal one.

The temperature evolution of the lattice parameters, resulting from the structural refinement, is displayed in Fig.~\ref{struc}(h). Table~\ref{atomic} further summarizes the refined structural parameter along with the Bragg agreement $R-$factors, whose low values highlight the goodness of the refinement model. Since the Cr atoms sit at the vertices of the unit cell, the evolution of the lattice parameters reflects directly the evolution of the in-plane Cr-Cr distances. This is a factor of major importance in the determination of the magnetic structure for 2D-TLAs \cite{rosenberg1982magnetic}. On cooling, the size of the crystallographic axes is reduced, due to thermal contraction, down to the magnetic transition temperature $T_{\rm N}$. As seen in Fig.~\ref{struc}(h), the contraction of the $c-$axis occurs at a much higher rate with respect to the $a-$ and $b-$axes, which is phenomenologically compatible with the 2D-TLA nature of the compound. At $T_{\rm N}=T_{\rm s}$ the first-order structural transition is clearly visible as a sharp change in the temperature dependence of the axes, with the major effect observed in the $ab-$plane. The synchronized nuclear and magnetic phase transition appearing at $T=30$~K indicated a strong spin-lattice coupling. Here, the ordering of the Cr magnetic moments seems to be the driving mechanism of a modification of the crystal lattice, aimed at stabilizing the competing coupling mechanisms simultaneously established among them. To further understand such synergic effect, a detailed understanding of the spin structure is required. A dedicated discussion around the structural evolution of the system in concomitance with the evolution of the magnetic structure will be presented in the next subsection.

\begin{table}[h!]
\renewcommand{\arraystretch}{1.25}
\small
  \caption{Lattice parameters, atomic positions, isotropic thermal displacement parameters and Bragg $R-$factors from refinement of neutron powder diffraction data on polycrystalline LiCrSe$_2$}
  \label{atomic}
  \begin{tabular*}{0.48\textwidth}{@{\extracolsep{\fill}}  c c c}
    \hline
                        & 300 K        & 2 K       \\
    \hline

    Space Group         & $P\bar{3}m1$  & $C2/m$  \\
    $a$ Å            & 3.65425(3)       & 6.32904(8)       \\
    $b$ Å            & 3.65425(3)       & 3.64287(5)      \\
    $c$ Å            & 6.28911(4)       & 6.25345(7)       \\
    $\alpha$        & 90$^{\circ}$     & 90$^{\circ}$       \\
    $\beta$         & 90$^{\circ}$     & 90.08(13)$^{\circ}$        \\
    $\gamma$        & 120$^{\circ}$    & 90$^{\circ}$       \\
    Li (x, y, z)    &  (0, 0, $\frac{1}{2}$) &  (0, 0, $\frac{1}{2}$) \\
    Cr (x, y, z)    &  (0, 0, 0) &  (0, 0, 0) \\
    Se (x, y, z)    & ($\frac{1}{3}$, $\frac{2}{3}$, 0.226(1))  &  (0.667(3), 0, 0.773(1)) \\
    B$_{Li}$  (Å$^2$) & 1.98(9)  & 0.85(7)  \\
    B$_{Cr}$  (Å$^2$) & 0.63(5)  &  0.23(3) \\
    B$_{Se}$  (Å$^2$) & 0.39(1)  &  0.018(7) \\
    $R_{B(nuc)}$ (\%)   & 4.2         & 4.5         \\
    \hline
  \end{tabular*}
\end{table}

\subsection*{NPD Results: Magnetic Refinement}
\begin{figure}[ht]
\centering
  \includegraphics[scale=0.5]{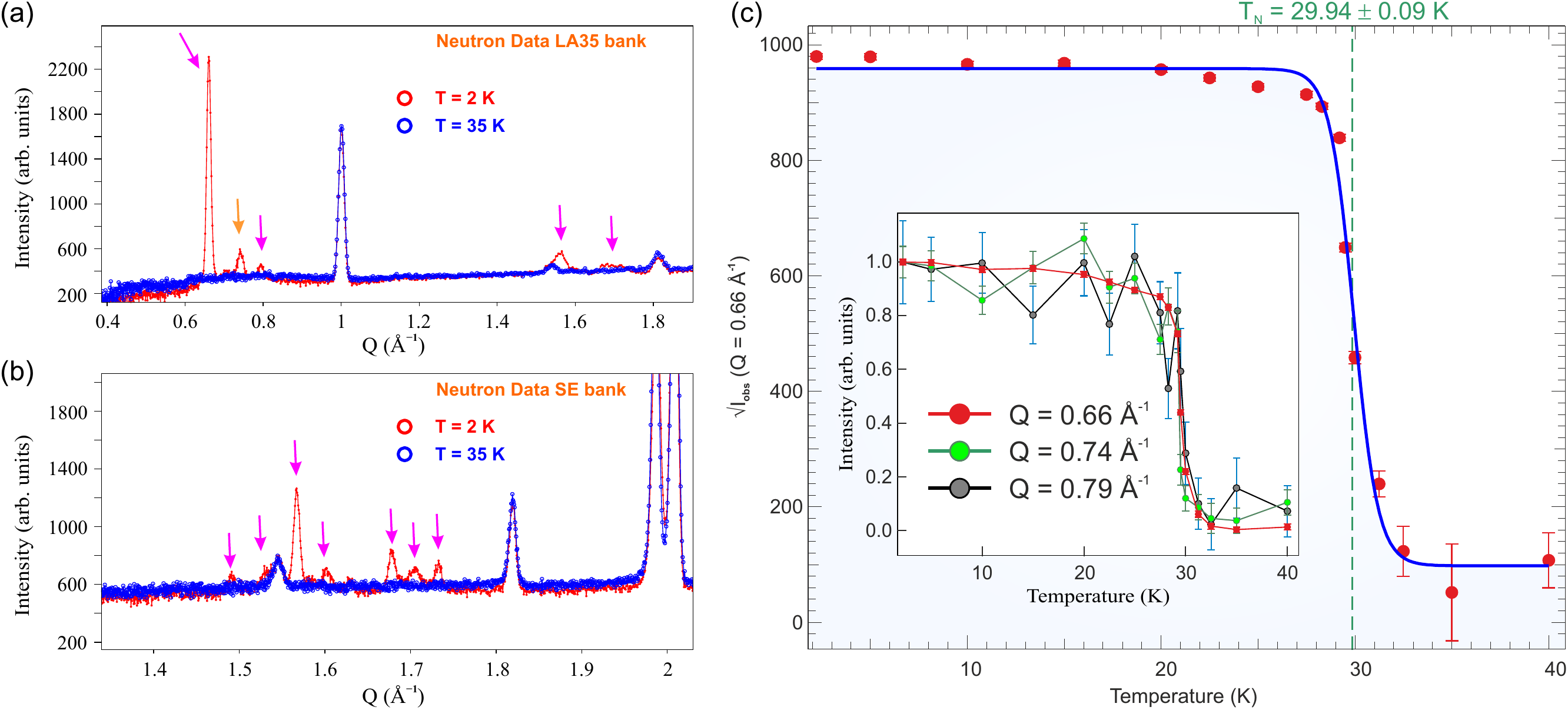}
  \caption{Neutron diffraction patterns (raw data) at $T=35$~K and $T=2$~K acquired using the (a) LA35 and (b) SE detector banks, respectively. The appearance of magnetic Bragg peaks below $T_{\rm N}\approx30$~K is evidenced by the magenta arrows (orange arrow indicates the strongest among the peaks un-indexed by the main magnetic phase, see text). (c) Square root of the integrated intensity of the strongest magnetic peak at $Q=0.66$~Å$^{-1}$, visible in the LA35 Bank. The continuous line is a fit to a sigmoid function for precise determination of the magnetic transition temperature. Inset: plot of the normalized integrated intensities of the three lower$-Q$ magnetic peaks.}
  \label{mag_peak}
\end{figure}

Several temperature dependent magnetic Bragg peaks appear below 30 K, as clearly seen in Fig.~\ref{split}(e,f). These peaks are assumed to be magnetic. Such transition is presented in higher detail in Fig.~\ref{mag_peak}(a,b), which displays a comparison between NPD patterns (SE $\rightarrow$ 90$^{\circ}$ and LA35 $\rightarrow$ 135$^{\circ}$ detector banks) acquired at $T=35$~K (just above $T_{\rm N}$) and at base temperature ($T=2$~K). The square root of the integrated intensity of the large magnetic peak in the lower $Q-$region ($Q=0.66$~Å$^{-1}$) is also plotted as a function of temperature [see Fig.~\ref{mag_peak}(c)]. The latter quantity reflects the order parameter of the magnetic transition which, as clearly seen from the plot in Fig.~\ref{mag_peak}(c), has a first-order character. The fit of the plot to a sigmoid function [solid line in Fig.~\ref{mag_peak}(c)], well follows the sharp profile of the data and provides the magnetic transition temperature $T_{\rm N}=29.94\pm0.09$~K (defined as the midpoint of the sigmoid curve). The inset of Fig.~\ref{mag_peak}(c) shows the normalized integrated intensities of the large magnetic peak together with the two neighboring small satellites, which all clearly manifest the same temperature dependent behavior. Nevertheless, as discussed further below, these peaks do not belong to the same magnetic phase. Here a distinction is necessary between a $main$ and a $secondary$ magnetic phase. In the immediate following, we will describe the main magnetic phase.

The magnetic propagation vector $q_{\rm main}$ was determined by a very extensive trial and error procedure with the software K-Search. As expected from our $\mu^+$SR data, it was found to be incommensurate: $q_{\rm main}$ = (0.045, $\approx \frac{1}{4}$, $\approx \frac{1}{2}$). Such propagation vector can successfully index the high intensity magnetic peak and most of the lower intensity ones as well. The remaining weak peaks were still un-indexed [e.g. at $Q=0.74$~Å$^{-1}$, orange arrow in Fig.~\ref{mag_peak}(a)], this issue will be addressed in the next section.

\begin{figure*}[ht]
\centering
  \includegraphics[keepaspectratio=true,width=175 mm]{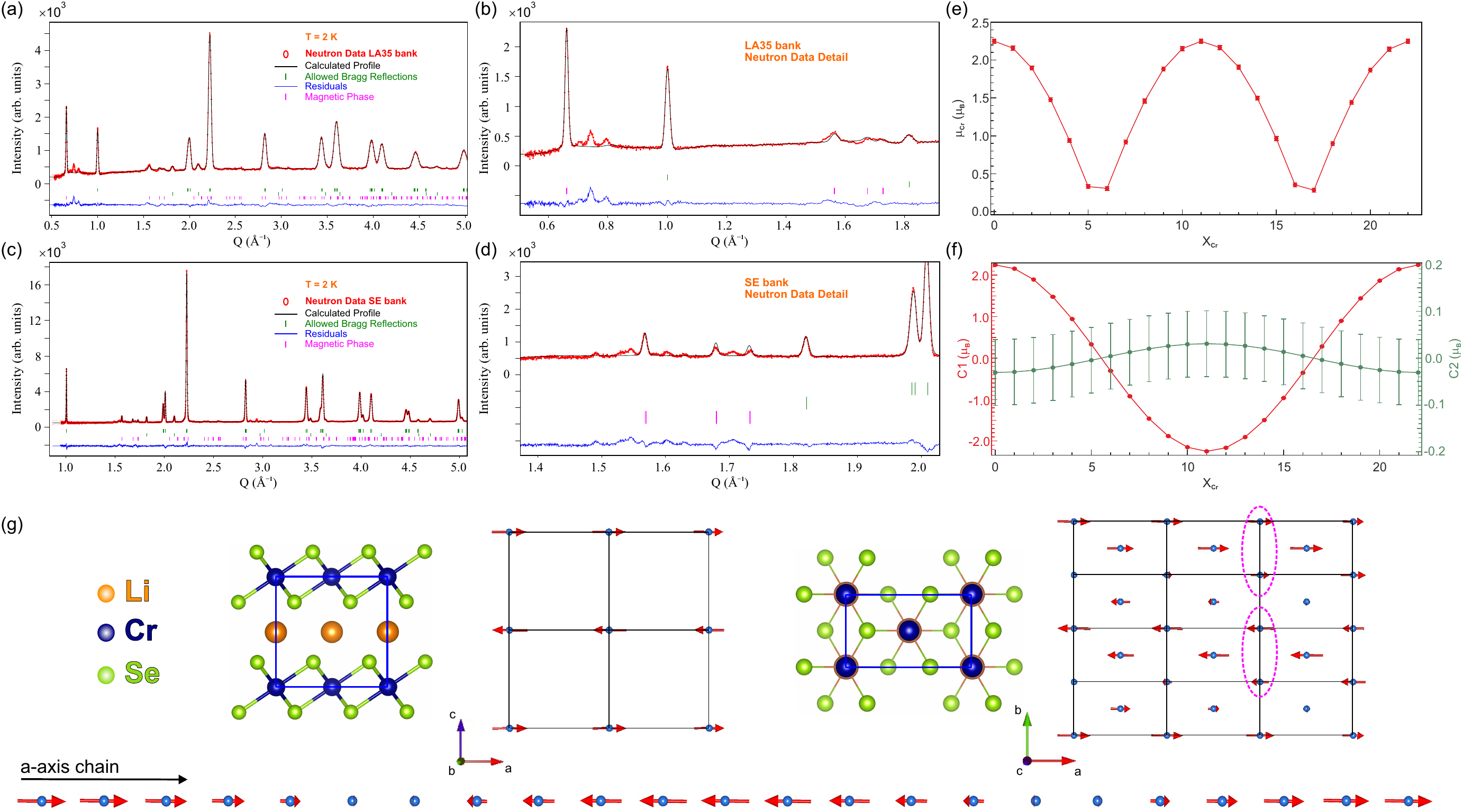}
  \caption{\textbf{Main Magnetic Phase}: Overview and close-up neutron diffraction pattern with the corresponding calculated pattern resulting from magnetic refinement at $T=2$~K using the (a,b) LA35 and (c,d) SE, detector banks, respectively. (e) Value of the spin only Cr moment and (f) the components of the axial spin vector C1 (left y-axis) and C2 (right y-axis). Both plotted along the $a-$axis as a function of the $x_{\rm Cr}$ atomic coordinate. (g) Graphic representation of the main magnetic structure in polycrystalline LiCrSe$_2$ oriented along the main axes of the unit cell (4 cells in the $ac-$plane, 12 cells in the $ab-$plane, and finally 22 cells along the $a-$axis at the very bottom). The up-up-down-down arrangement of the spins along the $b-$axis is highlighted by the dashed magenta ellipses. See also Table~\ref{table2} for refined parameters.}
  \label{q_dep_mom}
\end{figure*}

The LiCrSe$_2$ magnetic ($q_{\rm main}$) phase, refined by the Rietveld Method, includes a single magnetic Cr atom in its crystallographic site Cr(0 0 0). The scale factor and structural parameters were constrained to be equal to their counterparts in the nuclear LiCrSe$_2$ phase, for correct estimation of the Cr magnetic moment $\mu_{\rm Cr}$. The software BasIreps provided only one possible irreducible representation of the propagation vector group $G_k$ for $q_{\rm main}$, compatible with the space group $C2/m$ (IRrep $\Gamma_1$), consisting of three basis vectors BsV, with no imaginary components, parallel to the main crystallographic axes:

\begin{itemize}
\item BsV(1): Re (1 0 0), 
\item BsV(2): Re (0 1 0), 
\item BsV(3): Re (0 0 1). 
\end{itemize}

The refinement of the coefficients of the basis vectors C1, C2 and C3 provided a large component in the $a$ direction for the spin axial vector and a negligibly small component along the $b$ direction [Fig.~\ref{q_dep_mom}(a-g)].

Using IRrep $\Gamma_1$ in the magnetic structure refinement [shown in Fig.~\ref{q_dep_mom}(a-d)], it was possible to well capture the temperature dependent magnetic peaks of the diffraction pattern indexed by the k-vector $q_{\rm main}$. The resulting magnetic structure presents a very large magnetic unit cell, with AFM coupled layers stacked along the $c-$axis with a peculiar "breathing" arrangement of the moments within the $ab-$plane [Fig.~\ref{q_dep_mom}(e-g)]. The intra-layer orientation of the Cr moments along the $b-$axis is reminiscent of an alternate up-up-down-down sequence [see dashed magenta ellipses in Fig.~\ref{q_dep_mom}(g)], built up by double FM-coupled Cr spin chains running along the $a-$axis. This peculiar arrangement of the Cr moments is very similar to what was found for AuCrS$_2$ \cite{carlsson2011suppression}. However, in this case the orientation of the spins in the FM-coupled chains undergoes a change of direction, accompanied by a size modulation of the magnetic moment while moving along the $a-$axis [see plot (e) and bottom of panel (g) in Fig.~\ref{q_dep_mom}]. As a result, a periodic suppression of the magnetic moment appears along the chain. The modulation interval between two consecutive spin flips along the $a-$axis is 11 unit cells. Figure~\ref{q_dep_mom}(e,f) displays the calculated value of the Cr moment $\mu_{\rm Cr}$ as a function of the atomic coordinate $x_{\rm Cr}$. This is obtained as the modulus of the refined spin axial vector expressed in units of Bohr magneton. The periodic "breathing" feature of the magnetic moment modulation is clearly visible. The highest value for $\mu_{Cr}$ in the chain is 2.25(3)$\mu_B$, lower than the value previously estimated from bulk magnetic measurements $\approx$ 4 $\mu_B$ \cite{kobayashi2016competition}, which is closer to the ideal spin-only value $\mu_{3/2}$ = 2$\sqrt{S(S+1)}$ = 3.87 $\mu_B$ for an isolated Cr$^{3+}$ ion in the spin state $S=3/2$. However, in a covalently bonded system, a reduction in the value of the magnetic moment of a transition metal element with respect to the ideal value is expected, due to the competition between covalency and magnetism \cite{streltsov2016covalent}. Further, when geometric frustration is present, the appearance of staggered magnetic moments and a following reduction of $\mu_{\rm Cr}$ is very common \cite{Soubeyroux_1979,damay2011magnetoelastic,carlsson2011suppression,Schmidt_2017,Nozaki_2010,Matsubara_2020}. Therefore, in agreement with systems analogous to LiCrSe$_2$ \cite{carlsson2011suppression}, the value $\mu_{\rm Cr}$ = 2.25(3)$\mu_B$ estimated by NPD in the current work can be considered reliable, and consistent with a $S$ = 3/2 spin system scenario. The resulting parameters obtained from the refinement of the magnetic unit cell for the LiCrSe$_2$ magnetic phase are summarized in Table~\ref{table2}, together with their $R_{Bragg}$ agreement factors. The observed profiles are in very good agreement with the calculated models, as clearly seen in Fig.~\ref{q_dep_mom}(a-d). The goodness of the model is underlined by the value of the reliability $R-$factor for the magnetic phase, not exceeding a few percent. All the values reported in Tables \ref{atomic} and \ref{table2} are obtained from the global procedure involving the simultaneous Rietveld refinement of diffraction patterns from the 3 detector banks BS, SE and LA35.

\begin{table}[h]
\renewcommand{\arraystretch}{1.25}
\small
  \caption{\textbf{Main Magnetic Phase}: Magnetic refinement parameters and agreement factors for the neutron powder diffraction data from LiCrSe$_2$ at $T=2$~K}
  \label{table2}
  \begin{tabular*}{0.48\textwidth}{@{\extracolsep{\fill}}  c c }
    \hline
     Space Group         & $P1$  \\
    \hline
    \textit{a}$_m$ (\AA)   & 75.9486(1)    \\
    \textit{b}$_m$ (\AA)   & 14.57148(2)    \\
    \textit{c}$_m$ (\AA)   & 12.50691(2) \\
    $\alpha_m$             & 90$^{\circ}$    \\
    $\beta_m$              & 90.0762(13)$^{\circ}$  \\
    $\gamma_m$             & 90$^{\circ}$ \\
     \hline

    $R_{B(mag)}$ (\%)      &       2.91      \\
    $\mu_{Cr}$ ($\mu_B$)   &      2.25(3)      \\
    \hline
  \end{tabular*}
\end{table}

\section*{Discussion}
\begin{figure}[h!]
\centering
  \includegraphics[scale=0.41]{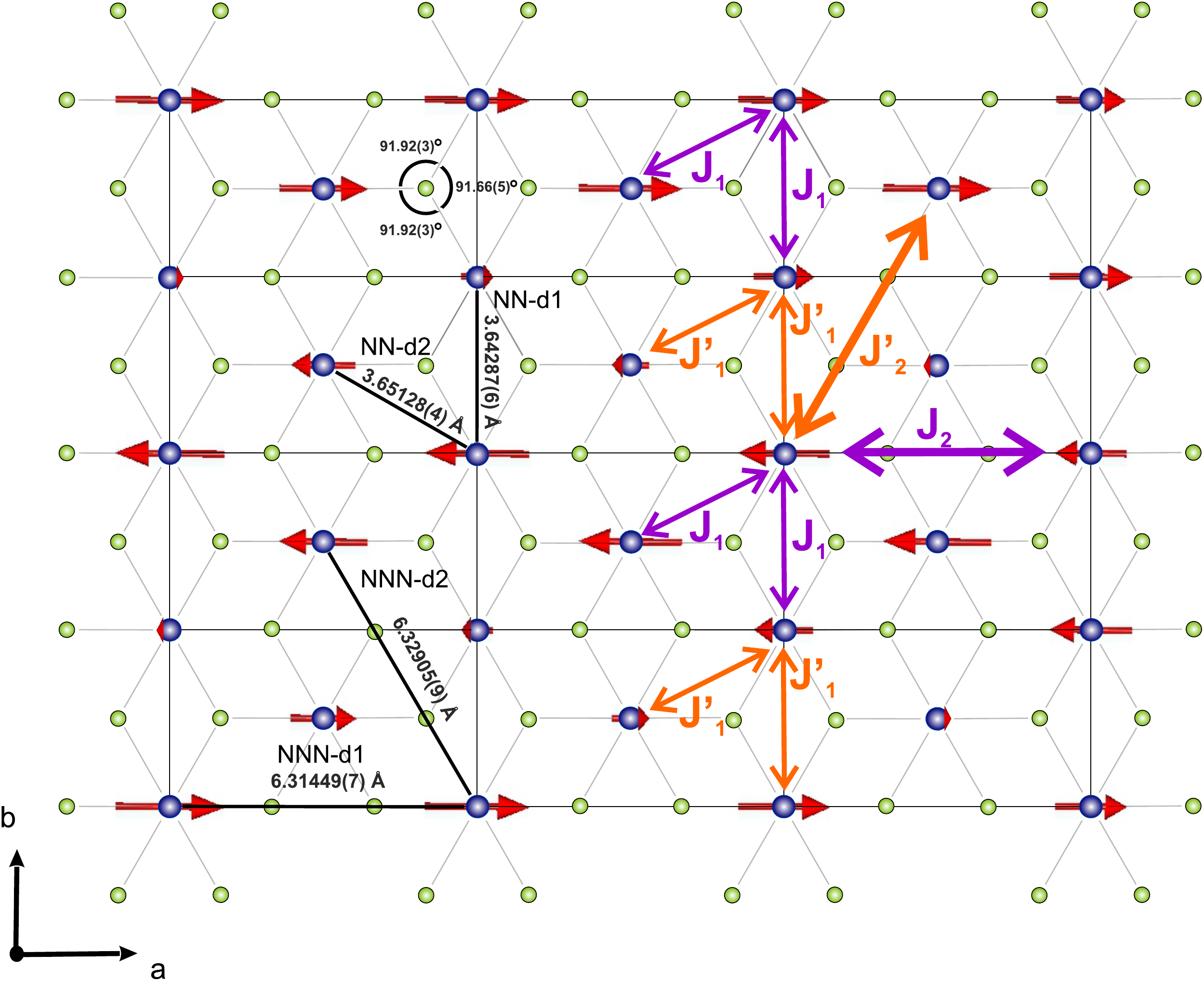}
  \caption{Pictorial representation of the magnetic couplings established between Cr$^{3+}$ atoms in the $ab-$plane of LiCrSe$_2$. The FM interactions $J_1$ and $J_2$ are displayed in orange while the AFM interactions $J'_1$ and $J'_2$ are displayed in purple. The Cr-Cr distances and angles established from the present diffraction studies are also shown (for the $C2/m$ space group at $T=2$~K, see also Table~\ref{dist}).}
  \label{mag_mech}
\end{figure}

The resulting incommensurate magnetic order, with itinerant frustration of the Cr moment \cite{campbell1995frustrated,li2019competing}, found in LiCrSe$_2$ is indeed rather peculiar. While the inter-planar magnetic ordering along the $c-$axis can be explained with an AFM coupling mediated by super-exchange interactions through the path Cr-Se-Se-Cr, to explain the complex intra-planar magnetic structure the coexistence of competing magnetic interactions must be considered [Fig.~\ref{mag_mech}]. In ternary chromium chalcogenide compounds, the magnetic interactions between two neighboring Cr$^{3+}$ ions can be of two types, as for the Goodenough-Kanamori-Anderson rules \cite{goodenough1955theory,kanamori1959superexchange,anderson1950antiferromagnetism}. Either a FM super-exchange interaction is established, by hybridization of the $p$ orbitals of an anion with the $d$ orbitals \footnote{Specifically with the half-filled $t_{2g}$ orbital of one Cr ion and the empty $e_g$ orbital of the other one} of the Cr ions forming a 90$^{\circ}$ bond of the type Cr-Anion-Cr; or an AFM direct exchange interaction is established between the half-filled $t_{2g}$ orbitals of two adjacent Cr$^{3+}$ ions. In the FM coupling, the Cr-Cr distance $d(Cr-Cr)$ is unimportant as long as the angle in the bond Cr-Anion-Cr is close to 90$^{\circ}$. On the other hand, for the AFM coupling, $d(Cr-Cr)$ has a major relevance. Rosenberg and coworkers identified a cutoff nearest neighbor (NN) Cr-Cr distance of $\approx$ 3.6 Å in the Cr layers of ternary chromium chalcogenides, for one mechanism to be dominant over the other \cite{rosenberg1982magnetic}. In particular, for $d(Cr-Cr)$ $<$ 3.6 Å the AFM interaction would be preferred, while for $d(Cr-Cr)$ $>$ 3.6 Å the FM interaction would be preferred. In LiCrSe$_2$ a dramatic structural change is observed, in concomitance to the magnetic transition at $T_{\rm N}$, leading to a modification of the Cr-Cr distances and of the angles $\widehat{Cr-Se-Cr}$, which are summarized in Table~\ref{dist}.

\begin{table}[h]
\renewcommand{\arraystretch}{1.25}
\small
  \caption{\ NN and NNN Cr-Cr distances and Cr-Se-Cr bonding angles just above and below the magnetic transition in LiCrSe$_2$}
  \label{dist}
  \begin{tabular*}{0.48\textwidth}{@{\extracolsep{\fill}}  c c c}
    \hline
                        & 35 K        & 2 K       \\
    \hline

    Space Group         & $P\bar{3}m1$  & $C2/m$  \\
    NN-$d_{1}(Cr-Cr)_{a-b}$ (Å)  & 3.65059(1)       & 3.64287(6)       \\
    NN-$d_{2}(Cr-Cr)_{a-b}$  (Å) & -       & 3.65128(4)      \\
    NNN-$d_{1}(Cr-Cr)_{a-b}$ (Å) & 6.32301(1)       & 6.31449(7)       \\
    NNN-$d_{2}(Cr-Cr)_{a-b}$  (Å) & -       & 6.32905(9)      \\
    $d(Cr-Cr)_{c}$       (Å)  & 6.25113(1)       & 6.25345(8)       \\
    $\widehat{Cr-Se-Cr}_1$     & 91.8863(1)$^{\circ}$     & 91.66(5)$^{\circ}$       \\
    $\widehat{Cr-Se-Cr}_2$        & 91.8863(1)$^{\circ}$     & 91.92(3)$^{\circ}$        \\
    \hline
  \end{tabular*}
\end{table}

In the trigonal phase just above the magnetic transition, the distance NN-$d(Cr-Cr)_{a-b}$ is very close to the critical value 3.6 Å and the angle $\widehat{Cr-Se-Cr}$ is very close to 90$^{\circ}$. In these conditions the AFM and FM couplings are equally likely to be established between NN Cr$^{3+}$ ions. As a consequence, the system is subjected to a geometrical frustration, which is removed through the energetically beneficial structural transition. This results in two non-equivalent NN and two non-equivalent next nearest neighbor (NNN) Cr$^{3+}$ distances, $d_{1}(Cr-Cr)_{a-b}$ and $d_{2}(Cr-Cr)_{a-b}$. In the low temperature monoclinic phase small distortions of the lattice occur, indeed both the NN distances are still very close to the critical value 3.6 Å (see Table \ref{dist}), and the angles $\widehat{Cr-Se-Cr}$ are still very close to 90$^{\circ}$. Nevertheless, as usually happens in frustration driven structural transitions, this distortion is enough to lift the degeneracy of the magnetic ground state. Here, a strong competition between AFM and FM interactions is established, which leads to an overall weakening of the exchange mechanism between NN Cr$^{3+}$ ions. At this point, the NNN Cr$^{3+}$ interactions must come into play [see Fig.~\ref{mag_mech}] to stabilize the magnetic structure. In particular, given the results of our magnetic refinement [Fig.~\ref{q_dep_mom}], we propose that a FM super-exchange interaction $J_2$ is always established through the Cr-Se-Se-Cr path between the NNN Cr$^{3+}$ atoms connected by the shorter distance NNN-$d_{1}(Cr-Cr)_{a-b}$. Further, the AFM super-exchange interaction $J'_2$ is always established through the Cr-Se-Se-Cr path between the NNN Cr$^{3+}$ atoms connected by the longer distance NNN-$d_{2}(Cr-Cr)_{a-b}$. Concerning the nearest neighbor interactions, a FM super-exchange interaction $J_1$ through the bond Cr-Se-Cr and an AFM direct exchange interaction $J'_1$ are established for both the NN-$d(Cr-Cr)_{a-b}$ distances, in an alternate arrangements between adjacent cells. This is not surprising since the geometric conditions for the occurrence of both couplings are fulfilled on both Cr-Cr paths, and there is no reason for the predominance of one mechanism over the other. As a result, these magnetic couplings, exemplified in Fig.~\ref{mag_mech}, are equiprobable on the same crystallographic sites. In addition, exchange interaction established among further than NNN Cr ions might be relevant in the determination of the ground state of LiCrSe$_2$. Indeed, it was found that in delafossite structures with S as chalcogenide ligand (e.g., AuCrS$_2$ and AgCrS$_2$), the NNN exchange interaction $J_2$ is typically comparable with the 3rd neighbor AFM exchange interaction $J_3$ \cite{ushakov2013magnetism}. Such mechanism of super-super-exchange, realized via two $p$-orbitals of the S ligands, seems to drive the magnetic transition and determine the magnetic ground state in these materials \cite{ushakov2013magnetism}. The third neighbor interaction was recently found to be relevant also for the determination of the magnetic ground state in the triangular Cr-oxide PdCrO$_2$ \cite{komleva2020unconventional}. The $p$-orbitals of the Se element have a a larger spatial extension with respect to the ones in S and O. Therefore, in qualitative agreement with the aforementioned systems \cite{ushakov2013magnetism,komleva2020unconventional}, $J_3$ could be expected to be even larger than $J_2$ in LiCrSe$_2$. However, first principle calculations would be needed to estimate the relevance of $J_3$ in the determination of the magnetic order on this material.

\begin{figure*}[ht]
\centering
  \includegraphics[scale=0.7]{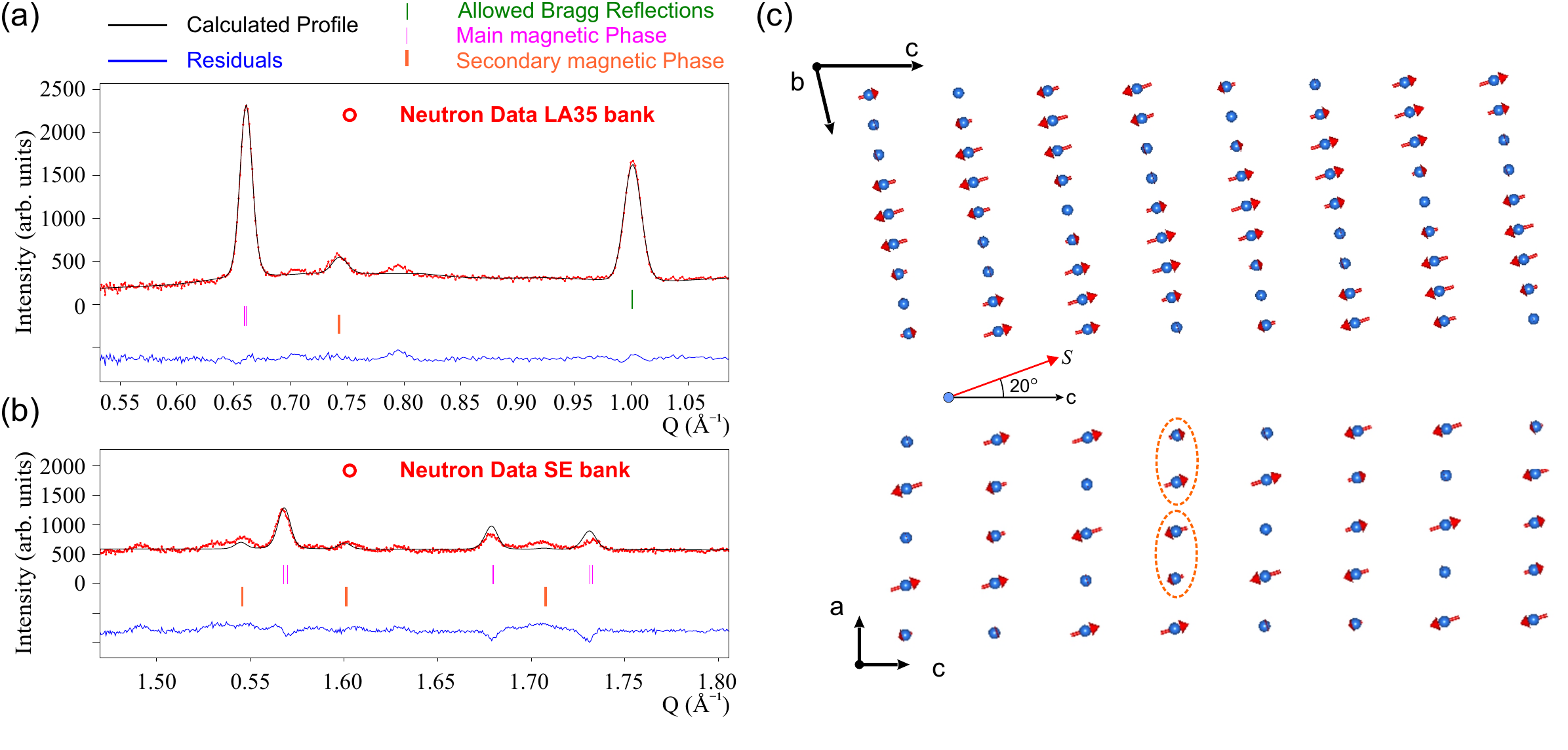}
  \caption{\textbf{Secondary Magnetic Phase:} Neutron diffraction pattern with the corresponding calculated pattern resulting from magnetic refinement at $T=2$~K for the two detector banks (a) LA35 and (b) SE, respectively. Both the main (magenta) and secondary (organge) magnetic phases are characterized by vertical tick bars of different color. (c) Pictorial representation of the spin structure for the secondary magnetic domain. See also Table~\ref{secondarytable} for refined parameters.}
  \label{secondary}
\end{figure*}

The magnetic structure induced by such competing interactions presents ferromagnetic double chains running along the $a-$axis, which are AFM coupled along the $b-$axis, similarly to AuCrS$_2$ and AgCrS$_2$ \cite{carlsson2011suppression,damay2011magnetoelastic}. The only difference is that for LiCrSe$_2$ the orientation of the Cr moment along the chains overturns periodically with a consequent modulation of its modulus. The alternating diagonal AFM-FM interactions between NN Cr$^{3+}$ atoms are probably responsible for this peculiar behavior. The proposed magnetic structure is consistent with the observation that magnetic correlations in compounds similar to LiCrSe$_2$ display significant further neighbor interactions involving Cr-Cr distances of 6–7 Å \cite{carlsson2011suppression,damay2011magnetoelastic}. Additionally, our results are in very good agreement with the findings of Kobayashi $et~al.$ \cite{kobayashi2016competition}, who observed an AFM character of the signal in magnetic susceptibility measurements. This suggested that further neighbor exchange interactions in LiCrSe$_2$ are in fact stronger than the nearest neighbor ones and of utmost importance in the determination of the magnetic ground state of LiCrSe$_2$. In reference \cite{kobayashi2016competition} it was also hypothesized that LiCrSe$_2$ is likely to possess a magnetic ground state different from other triangular lattice Cr chalcogenides with a similar structure and behavior \cite{carlsson2011suppression,damay2011magnetoelastic,rasch2009magnetoelastic}. The results of our work confirm this conjecture and provide a qualitative explanation for the difference between LiCrSe$_2$ and the other systems: in AuCrS$_2$ and AgCrS$_2$ and CuCrS$_2$ the in-plane Cr-Cr distances are always 3-4$\%$ different from the critical value found by Rosenberg. On the other hand, for LiCrSe$_2$ the in-plane Cr-Cr distances are just $\approx$ 1$\%$ different from the critical value 3.6 Å. The Cr-Anion-Cr angles are very close to 90$^{\circ}$ in all cases. Therefore, in LiCrSe$_2$ distinctive geometric conditions are realized, which lead to a configuration of the overlapping electronic orbitals that allow the conflictual coexistence of FM and AFM exchange interactions over the same crystallographic sites, making this material unique in the 2D-TLA family. Future modeling efforts would be very beneficial to systematically understand the complex spin interactions and order in LiCrSe$_2$. The detailed information presented in this work, concerning the atomic/spin arrangement across the transition, will most likely be a valuable input for such modeling efforts.

Going back to the un-indexed weak Bragg peaks found in the low-temperature $C2/m$ phase, it is important to emphasize that we clearly do not believe that these peaks are from an impurity phase. This is because there is no trace of magnetic impurities in the room temperature diffraction patterns, there is no sign of phase separation in the wTF $\mu^+$SR data, and the additional peaks appear exactly at the same structural/magnetic transition temperature of the LiCrSe$_2$ main phase ($T_{\rm s}=T_{\rm N}=30$~K). In our refinement efforts, a huge number of different magnetic propagation vectors in combination with different crystal structures and super-cells, including higher-order harmonics of the parent cell, were attempted for the purpose of simultaneous indexing of all the peaks in the diffraction pattern. However, these efforts did not produce any outcome and, even if the $C2/m$ nuclear structure combined with magnetic propagation vector $q_{\rm main}$ gives a very nice result, a few weak Bragg peaks were still left un-indexed [e.g. $Q=0.74$~Å$^{-1}$ being the strongest un-indexed peak, see Fig.~\ref{mag_peak}(a) and Fig.~\ref{q_dep_mom}(b)]. Considering the coinciding structural and magnetic phase transition occuring at $T\approx30$~K, it is indeed challenging to identify whether new Bragg peaks appearing at low temperature have nuclear or magnetic origin. In this work we therefore combined NPD with synchrotron XRD experiments in order to improve the identification process. Indeed, the low-temperature peak at $Q=0.74$~Å$^{-1}$ is not visible in the $T=11$~K XRD data and would then be assumed to have magnetic origin (but we will get back to this issue). Considering that the ZF $\mu^+$SR results indicated the presence of two slightly different magnetic phases we decided to try to index the few weak remaining low-temperature peaks with a secondary magnetic phase. Considering the fact that the ZF asymmetry of the the first Bessel component, $A_{\rm Bess1}$, was so much smaller than $A_{\rm Bess2}$, it seems reasonable to assume that the weak un-indexed magnetic Bragg peaks could be associated with $A_{\rm Bess1}$.

To deduce the secondary magnetic phase (consistent with the presence of a magnetic supercell) we utilized a similar procedure as for the identification of the main magnetic structure. As a result we find an incommensurate propagation vector $q_{\rm secondary}$ = (0.7249(5), -0.0885(3), -0.1292(9)) and only one possible irreducible representation of the propagation vector group $G_k$ for $q_{\rm secondary}$, compatible with the space group $C2/m$. Again, this IRrep consists of three basis vectors, with no imaginary components, parallel to the main crystallographic axes, BsV(1): Re (1 0 0), BsV(2): Re (0 1 0), BsV(3): Re (0 0 1) with coefficients (0.45(11), 0, 1.26(5)) and a canting angle of $\sim 20^{\circ}$ from the $c-$axis for the spin axial vector. The resulting secondary magnetic structure [refinement in Fig.~\ref{secondary}(a,b)] appears similar yet complementary to the main structure; here the canted modulated FM chains run along the $c-$axis while the up-up-down-down arrangement of the Cr spins develops along the $a-$axis [Fig.~\ref{secondary}(c)]. The refined model inclusive of this additional magnetic phase is able to well capture most of the smaller magnetic satellites in the experimental data [orange tick-marks in Fig.~\ref{secondary}(a,b)]. The ordered Cr moment of the secondary domain is refined as $\mu_{Cr}=1.33(2)\mu_B$. The resulting parameters obtained from refinement of the secondary magnetic unit cell are summarized in Table~\ref{secondarytable}.

\begin{figure}[ht]
\centering
  \includegraphics[scale=1]{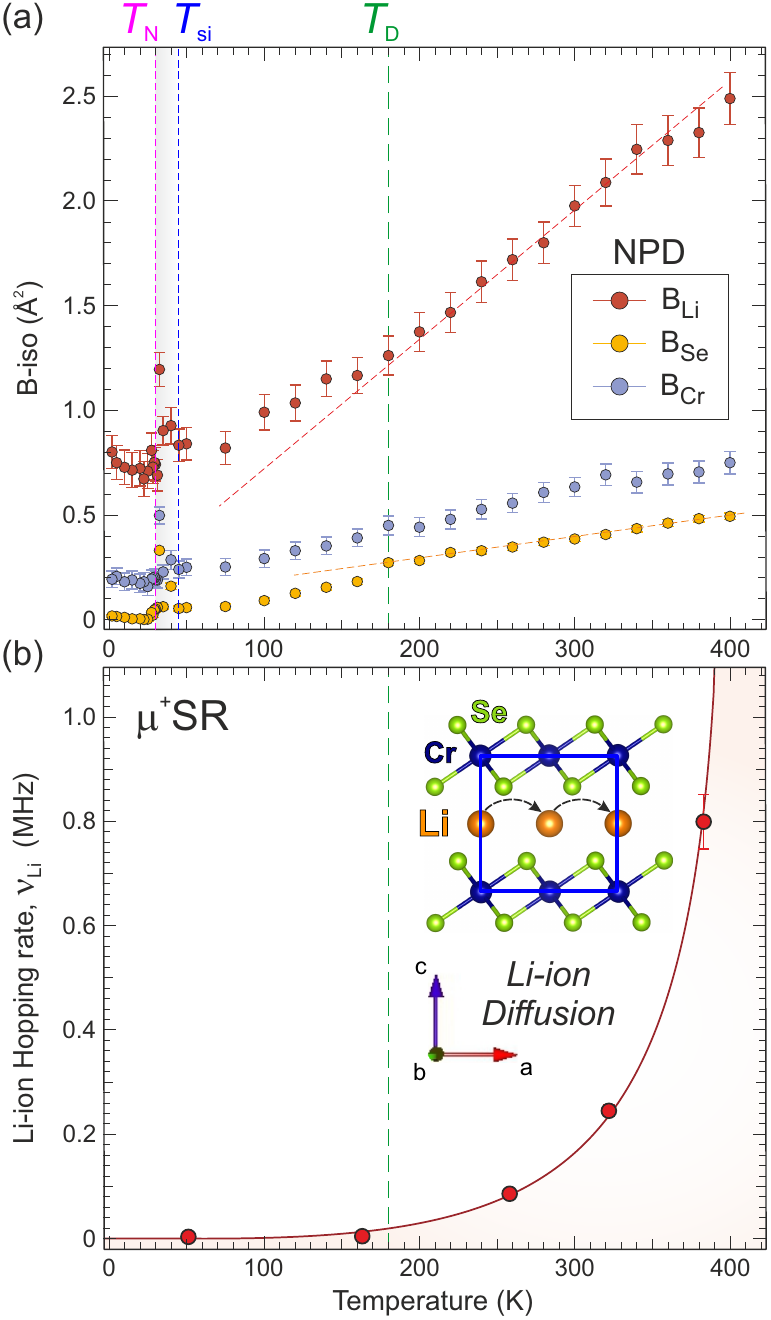}
  \caption{Temperature dependence of (a) $B_{\rm iso}$ parameters from NPD data, and (b) Li-ion hopping rate ($\nu_{\rm Li}$) from $\mu^+$SR measurements. Inset in (b) show the layered structure where Li-ions diffuse (hop) from site to site above $T_{\rm D}\approx180$~K. $T_{\rm si}\approx45$~K indicate the onset of structural instability. Solid and dashed lines are guides to the eye. }
  \label{Lidynamics}
\end{figure}

\begin{table}[h]
\renewcommand{\arraystretch}{1.25}
\small
  \caption{\textbf{Secondary Magnetic Phase}: Magnetic refinement parameters and agreement factors for the neutron powder diffraction data from LiCrSe$_2$ at $T=2$~K}
  \label{secondarytable}
  \begin{tabular*}{0.48\textwidth}{@{\extracolsep{\fill}}  c c }
    \hline
     Space Group         & $P1$  \\
    \hline
    \textit{a}$_m$ (\AA)   & 6.32905(1)    \\
    \textit{b}$_m$ (\AA)   & 40.0716(2)    \\
    \textit{c}$_m$ (\AA)   & 50.0276(2) \\
    $\alpha_m$             & 90$^{\circ}$    \\
    $\beta_m$              & 90.0762(13)$^{\circ}$  \\
    $\gamma_m$             & 90$^{\circ}$ \\
     \hline

    $R_{B(mag)}$ (\%)      &       25.7     \\
    $\mu_{Cr}$ ($\mu_B$)   &      1.33(2)      \\
    \hline
  \end{tabular*}
\end{table}

Comparing the magnetic peaks visible in the LA35 bank [see Fig.~\ref{secondary}(a)], the magnitude of the integrated intensity of the largest peak belonging to the main magnetic phase (corresponding to the (0 0 1) and (0 0 0) reflections) is a factor of $\sim8$ larger than the integrated intensity of the largest peak belonging to the secondary phase (corresponding to the (0 0 0) reflection). This ratio is roughly consistent with the calculated values of the ordered moments in the two phases, since the integrated intensity of the magnetic Bragg peak is proportional to $\mu_{Cr}^2$. The presence of two distinct incommensurate magnetic phases is also consistent with the results of the ZF $\mu^{+}$SR analysis. Here it was found that the volume fractions of the two incommensurate components were identified as $A_{\rm Bess2}\approx12\cdot A_{\rm Bess1}$, which is compatible with the refinement of the main:secondary phases. Moreover, the clear canting angle found in the secondary magnetic phase is suggestive of the presence of asymmetric magnetic coupling, possibly due to the presence of Dzyaloshinskii-Moriya interaction (DMI). This would be consistent with a recent theoretical work  \cite{li2022large}, which indeed predicts the occurrence of large DMI. Nevertheless, the rather large value of the reliability $R-$factor for the secondary magnetic phase ($R_{\rm B(mag)}^{\rm secondary}=25.7\%$) makes this interpretation a bit uncertain. Finally, we speculate about the electronic origins of the magnetic peaks and the very unusual spin arrangement realized in LiCrSe$_2$. The spin density modulation along the FM chains and the incommensurate positions of the magnetic peaks might be originated by a combination of localized and itinerant moments, with a phenomenology similar to both the Na$_x$CoO$_2$ compound \cite{Julien_2008} (see also below) as well as the intermetallic material ErPd$_2$Si$_2$. The latter in fact also manifests two distinct incommensurate spin states \cite{li2015distinct}. Looking at the size of the magnetic peaks belonging to the two structures, it can be argued that the moments for $q_{\rm main}$ have mostly a localized character, while $q_{\rm secondary}$ are mainly itinerant. Although being consistent with the experimental observations, single crystal magnetic field dependent studies would be necessary to confirm or deny such hypothesis.

Another option is that the small un-indexed low-temperature Bragg peaks are in fact not magnetic but instead have nuclear origin related to charge ordering. Indeed, we could consider the appearance of a superstructure induced by e.g. Li(-vacancy) ordering. In principle such superstructure peaks should be visible also in the XRD, however, if they are indeed originating from the light Li atoms (low $Z$) the peaks might be difficult to discern (or even invisible) for X-rays. If we look more closely into the structural refinement from NPD, and specifically on the isotropic thermal displacement parameters $B_{\rm iso}$ shown in Fig.~\ref{Lidynamics}(a) there are some indications. First of all, the structural transition at $T_{\rm s}=30$~K is clearly visible. It is also clear that $B_{\rm iso}$ for all atoms actually deviate from the general trend already above the transition, indicating the onset of a structural instability (si) at around $T_{\rm si}\approx45$~K. This would actually explain the broad transition found in the wTF $\mu^+$SR data [Fig.~\ref{muon}(b)], just above the first order downturn. It is also clear that the Li atoms undergo the highest thermal displacement (cf. Cr and Se). At high temperature $B_{\rm Li}$ display an almost linear temperature dependence. However, below approximately 180~K a clear deviation appears. If we consider the similarity of LiCrSe$_2$ with several other well-known layered TMO compounds (e.g. LiCoO$_2$ \cite{Mizushima_1980}), and their utilization in rechargeable batteries \cite{Xie_2020}, we should consider the possibility of, and effects from, Li-ion dynamics. In fact our team has developed and pioneered an experimental method to investigate ion dynamics by $\mu^+$SR. A detailed description of the method and selected examples can be found in Refs.~\cite{Sugiyama_2009_diff,Mansson_2013,Muon_book_new}. Briefly, the muon spin is so sensitive to internal fields that it is capable of even probing static and dynamic changes in the nuclear moment. When implanted in these compounds in their higher temperature paramagnetic (PM) phase, the muon will mainly 'feel' the nuclear moment of the alkali metal (here Li), measured in both ZF and applied magnetic field parallel to the initial direction of the muon spin polarization (LF-$\mu+^+$SR). If the Li-ion is static, only a static nuclear field distribution ($\Delta$) is probed. However, if Li-ion dynamics sets in, an additional field fluctuation rate ($\nu$) component will be present. In the simplest interpretation, $\nu$ would directly translate into a Li-ion hopping rate that can be directly probed as a function of, e.g. temperature. In fact our team recently conducted initial test measurements (still to be completed) of Li-ion dynamics in LiCrSe$_2$ and $\nu(T)$ was successfully extracted at a few selected temperatures. As shown in Fig.~\ref{Lidynamics}(b), $\nu(T)$ show a clear exponential increase starting around $T_{\rm D}\approx180$~K, typical for a thermally activated diffusion process. This means that the Li-ions are fully mobile at least down to $T_{\rm D}\approx180$~K, which would indicate the possibility for Li order to appear during cool-down of the sample. This would clearly also explain the anomaly found for $B_{\rm Li}$ at $T_{\rm D}$. In fact a smaller kink is also visible in $B_{\rm Se}$, which is reasonable considering that the Se sits next to the Li layers in the LiCrSe$_2$ structure [see inset of Fig.~\ref{Lidynamics}(b)]. Interestingly, this behavior is different with respect to the one observed in the directly related compound LiCrO$_2$, which was proven to be electrochemically inactive up to 480 K, despite its striking structural similarity with LiCrSe$_2$. Therefore, the diffusion properties of this material could provide a hint on how to increment the Li-ion mobility in layered structures.

In fact, the dynamic phenomenology manifested by LiCrSe$_2$ is reminiscent of the behavior of the rather well-known and strongly related Na$_{\rm x}$CoO$_2$ (NCO) compound \cite{Sugiyama_2003,Hertz_2008,Schulze_2008,Medarde_2013,Sassa_2018}. From very extensive studies of NCO it has been shown that Na-ions (vacancies) can order in many different types of configurations \cite{Roger_2007,Morris_2009,Meng_2008}. It was also clearly shown that the Na-ion are in fact mobile even below room temperature and that the dynamics allow the Na-ions to arrange themselves in different patterns at lower temperatures \cite{Weller_2009,Willis_2018,Medarde_2013}. Even in the case of NCO with the heavier Na-ions, it was not straightforward to identify the superstructure ordering. Several of the successful studies were performed using neutron scattering and, in more or less all cases, the superstructures could only be identified using single crystalline samples rather than powder. It was also in single crystals where the most recent study of the in-plane as well as out-of-plane Na ordering was studied, this time using XRD \cite{Galeski_2016}. In the case of LiCrSe$_2$ it is therefore not improbable that the small satellites found in NPD could be nuclear superstructures and still not visible in the XRD pattern.

The comparison with NCO actually opens up an even more interesting possibility. Considering that both LiCrSe$_2$ and NCO have their magnetic ions sitting on the triangular (frustrated) lattice with competing magnetic interactions, the spins are highly susceptible to any additional perturbation. In the case of NCO it was clearly shown that the Na-ion ordering actually creates a periodic coulomb potential (charge) landscape that is able to tune the electronic states and magnetic properties within the TMO layers \cite{Julien_2008}. It has even been proven that, by controlling the cooling rate of the sample, thereby gaining control of dynamics and vacancy order, different magnetic ground states could be achieved for the NCO compound \cite{Schulze_2008,Galeski_2016}. Further, by electrochemical cycling of NCO inside a battery cell (i.e. changing the Na content $x$ for Na$_x$CoO$_2$), very detailed studies could be performed on the nuclear structure \cite{Berthelot_2010} as well as how this strongly affects the magnetic phase diagram \cite{Foo_2004,Sassa_2018}. The conditions for such phenomenology could in fact be verified also for LiCrSe$_2$. We could imagine that if Li-ions organize themselves in an ordered pattern, their periodic potential could affect the spin order within the triangular lattice Cr planes. Such effect could be the origin for the peculiar main magnetic phase ($q_{\rm main}$) firmly established in LiCrSe$_2$, as well as for the potential secondary magnetic domain ($q_{\rm secondary}$). Nevertheless, NCO and LiCrSe$_2$ present some differences, in NCO there are two Na sites and neither is fully occupied, while in LiCrSe$_2$ there is only one Li site. Additionally, we found in our refinement that all the atomic sites in our sample are fully occupied (i.e. no vacancies). On the other hand, the refined magnetic (main) phase has a very long periodicity (large unit cell), which would fit well with a Li-vacancy order based on very few existing vacancies that we did not discern in our refinement. Finally, it is of course possible to have Li order without vacancies, e.g. in the form of a charge-density-wave (CDW). Future more detailed studies of Li-ion ordering and dynamics in LiCrSe$_2$ coupled to the magnetic properties would clearly be very interesting.

In summary, we present a few possible scenarios attempting explain the un-indexed Bragg peaks found in our NPD data. To ultimately discern if the additional peaks are of nuclear or magnetic origin, single crystal polarized neutron diffraction would be the optimal approach. However, to the best of our knowledge no large enough LiCrSe$_2$ single crystals are presently available.

\section*{\label{sec:discussion}Conclusions}
The magnetic and crystal structure of the triangular antiferromanget LiCrSe$_2$ has been solved at high and low temperature by means of neutron diffraction, synchrotron X-ray diffraction and muon spin rotation experimental techniques. The unique arrangement of the Cr atoms in this material allows the simultaneous presence of competing nearest neighbor ferromagnetic and antiferromagnetic exchange couplings over the same crystallographic sites, leading to a predominance of the next nearest neighbor interactions (which are also both FM and AFM) in the determination of the magnetic ground state in LiCrSe$_2$. As a result, a structural transition from trigonal $P\bar{3}m1$  to monoclinic $C2/m$ crystal system induced by strong magnetoelastic coupling occurs, accompanied by the formation of two up-up-down-down magnetic structures (a main one and a secondary one) with itinerant modulation of the Cr spin axial vector, in direction and modulus, and incommensurate $q-$vectors $q_{main}$ = (0.045, $\approx \frac{1}{4}$, $\approx \frac{1}{2}$) and $q_{secondary}$ = (0.7249(5), -0.0885(3), -0.1292(9)). Such arrangement of the Cr spins is reminiscent of the magnetic ordering established in the Cr sulfide triangular antiferromagnets AuCrS$_2$ and AgCrS$_2$ which also manifest strong magnetoelastic coupling. However, the unprecedented geometric conditions realized in LiCrSe$_2$ provide access to a unique combination for the competing magnetic mechanisms, resulting in a much more complex magnetic ground state in which the magnetic moment is periodically suppressed and changes direction. This peculiar behavior might be originated by a dual itinerant/localized nature of the electrons in this system. From the results of this work LiCrSe$_2$ was proven to be an extremely fascinating study case, which lends itself to very promising future single crystal investigations, and constitutes an important step forward in understanding the spin-lattice relationship in strongly correlated spin systems.

\section*{Methods}

Polycrystalline LiCrSe$_2$ was prepared by direct reaction of the Li, Cr, and Se elements. Further details on the sample preparation can be found in reference \cite{kobayashi2016competition}.

The low-temperature $\mu^+$SR spectra have been acquired, as a function of temperature, at the multi purpose muon beam-line M20, at the Canada's particle accelerator centre TRIUMF. $\sim$ 200 mg of sample in powder was packed in a 1$\times$1 cm$^2$ area Aluminum-coated Mylar tape (0.05 mm thickness), in order to reduce the background signal. This envelope was attached to a low background sample holder inserted in a helium exchange gas cryostat  (temperature range: 2 K to 300 K).

The high-temperature $\mu^+$SR data were obtained using the surface muon beamline S1 at the high intensity proton accelerator facility J-PARC, in Japan. The LiCrSe$_2$ powder was pressed into a pellet introduced into hermetically sealed non-magnetic titanium cells. The surface of the pellet facing the muon beam was covered using a very thin (50~$\mu$m) titanium window. By using a closed-cycle refrigerator (CCR), data was collected in the temperature range $T=50-375$~K. A more detailed description of the sample cell and experimental setup can be found in Ref.~\cite{Matsubara_2020}.

The XRD measurements were collected at the instrument I11 of the Diamond Light Source in Didcot (UK). The sample was mounted by filling and sealing an Al capillary with internal diameter 0.7 mm, which was in turn attached to a phoenix cryostat, with a 10$^{\circ}$ sample rocking stage, which allows measurements in a temperature range from 11 K to 300 K.

The neutron diffraction measurements were performed at the time-of-flight (ToF) powder diffractometer iMATERIA \cite{ishigaki2009ibaraki}, at J-PARC, in Japan. The powder samples ($\sim$ 500 mg) were mounted into cylindrical vanadium cells with diameters and 5 mm and sealed with an aluminium cap, aluminium screws and indium wire. The cell was mounted on a closed cycle refrigerator to reach temperatures between 2 K and 300 K. iMATERIA allows for the simultaneous adoption of different detector banks for wide d-range ($Q-$range) coverage. The backscattering bank (BS), suitable for detailed structural characterization, has a resolution $\Delta d / d$ = 0.16$\%$ and allows a d-range from 0.181 Å up to 5.09 Å, the 90-degree bank (SE) has a resolution $\Delta d / d$ = 0.5$\%$ and allows a d-range from 0.255 Å up to 7.2 Å, and the low angle bank (LA35), ideal for the detection of low $Q-$range magnetic Bragg peaks, has a d-range from 0.25 Å up to 40 Å.

Due to high sensitivity of the sample to air and moisture, the sample preparations for all the measurements were carried out in a controlled environment using Ar and/or He glove-boxes.

The crystal and magnetic structure determination was carried out with the help of the Bilbao crystallographic server \cite{aroyo2006bilbao1,aroyo2006bilbao2}, and the FullProf software suite \cite{rodriguez1993recent} was employed for the data refinement. All images involving crystal structure were made with the VESTA software \cite{momma}, the parameter plots and fitting are produced with the software Igorpro \cite{igor} and the $\mu^+$SR data were fitted using the software package \textit{musrfit} \cite{musrfit}.

\bibliography{Refs}

\section*{Acknowledgements}

The $\mu^+$SR measurements were performed at the instrument M20 of the muon source TRIUMF (beamtime proposal: M1673) as well as at the S1 beamline of J-PARC (beamtime proposal: 2019A0327). The XRD measurements were performed at the instrument I11 of the synchrotron facility Diamond (beamtime proposal: CY23840). The NPD measurements were performed at the instrument iMATERIA of the neutron spallation source J-PARC (beamtime proposal: 20180056A).
The authors wish to thank Hiroshi Nozaki for his support during the $\mu^+$SR experiment, and Juan Rodriguez-Carvajal for the assistance he provided on the neutron diffraction data refinement, along wit the staff of TRIUMF, Diamond and J-PARC for the help in the experimental measurements.
This research is funded by the Swedish Foundation for Strategic Research (SSF) within the Swedish national graduate school in neutron scattering (SwedNess), as well as the Swedish Research Council VR (Dnr. 2021-06157 and Dnr. 2017-05078), and the Carl Tryggers Foundation for Scientific Research (CTS-18:272). J.S. is supported by the Japan Society for the Promotion Science (JSPS) KAKENHI Grant No. JP18H01863 and JP20K21149. Y.S. and O.K.F. are funded by the Chalmers Area of Advance - Materials Science. S.K. and K.Y. are supported by JSPS KAKENHI Grant No.18KK0150. The work at the University of Zürich and the University of Geneva was supported by the Swiss National Science Foundation under Grant No. PCEFP2-194183.

\section*{Author contributions statement}

E.N., J.S., and M.M. conceived the experiments. E.N., O.K.F., N.M., C.T., T.M., A.O., A.K., I.U., Y.S., J.S. and M.M. conducted the experiments. E.N, J.S., N.M., O.K.F., J.H.B., V.P, and M.M. analyzed the results. The samples were synthesized by S.K. and K.Y., as well as by C.W. and F.O.v.R.; they also conducted the initial sample characterizations. E.N and M.M. made all the figures. E.N. created the first draft, and all co-authors reviewed and revised the manuscript. 

\section*{Data availability statement}

All the data of this work are available from the corresponding authors on request.

\textbf{Competing interests} 

The authors declare no competing interests.  

The corresponding author is responsible for submitting a \href{http://www.nature.com/srep/policies/index.html#competing}{competing interests statement} on behalf of all authors of the paper.
\end{document}